\documentclass[10pt]{article}

\usepackage[dvips]{graphicx}
\usepackage{amssymb}

\newcommand{\eps} {\varepsilon}

\begin{document}

\title{Ordering mechanisms in confined diblock copolymers}
\author{Yoav Tsori \thanks{ESPCI, 10 Rue Vaquelin, Paris
75231 Cedex 05, France} \ and David Andelman\thanks{School of
Physics and Astronomy, Raymond and Beverly Sackler Faculty of
Exact Sciences, Tel Aviv University, Ramat Aviv, Tel Aviv 69978
Israel. E-mail: andelman@tau.ac.il}}

\date{16/7/2002}

\maketitle
\begin{abstract}
We present several ordering mechanisms in diblock copolymers. For
temperatures above the order-disorder temperature and in the weak
segregation regime, a linear response theory is presented which
gives the polymer density in the vicinity of confining flat
surfaces. The surfaces are chemically patterned where different
regions attract different parts of the copolymer chain. The
surface pattern or template is decomposed into its Fourier modes,
and the decay of these modes is analyzed. The persistence of the
surface pattern into the disordered bulk is given for several
types of patterns (e.g. uniform and striped surface). It further
is shown that complex morphology can be induced in a thin film
even though the bulk is disordered. We next consider lamellar
diblock copolymers (low temperature regime) in the presence of a
striped surface. It is shown that lamellae will appear tilted
with
respect to the surface, if the surface periodicity is larger than
the bulk one. The lamellae close to the surface are strongly
distorted from their perfect shape. When the surface and lamellar
periodicities are equal lamellae form perpendicular to the
surface. Lastly, the transition from parallel to perpendicular
lamellae to the surfaces in a thin film is presented. The
transition between the two states depends on the surface
separation and strength of surface interactions. We further
calculate the phase diagram in the presence of perpendicular
electric field favoring perpendicular ordering. In the strong
segregation limit we introduce a simple model to calculate the
phase diagram of the fully parallel, fully perpendicular and
mixed
(parallel and perpendicular) states.

\end{abstract}

{\bf  keywords: block copolymers, lamellar ordering, electric
fields.}


\newpage
\section{Introduction}

Block copolymers (BCP) are polymeric systems where each polymer
chain is composed of several chemically distinct homopolymer
blocks,  connected together by a covalent bond.  At high
temperatures, BCP have a disordered phase, while at low
temperatures, the macroscopic phase separation is hindered
because
the two (or more) immiscible sub-chains cannot be detached from
each other as they try to phase separate. Hence, BCP phase
separate into a variety of micro-ordered structures, with
characteristic size which depends on the BCP chain length
 and other system parameters \cite{hamley1}. The morphology and
structure of the
prevailing phase depends on the lengths of constituent sub-chains
(also called blocks), the chemical interactions between the
blocks, the temperature and the chain architecture. The BCP
micro-domain size ranges from about $10$ to several hundreds
nanometers. This fundamental periodicity should be distinguished
from the macro-domain size of several micrometers where one
ordered phase (e.g., a lamellar phase) breaks into many domains
(or grains) each having of a different orientation and separated
by grain boundaries.

 Block copolymers can be viewed as composite materials from
the mechanical point of view \cite{yachin1}. By connecting
together a stiff (rod-like) block with a flexible (coil) block,
one can obtain a material which is rigid, but not brittle
\cite{yachin2}. Moreover, the interplay between flexibility and
toughness can be controlled by temperature. Different chain
architecture (ring or star-like) may lead to novel mechanical and
flow properties \cite{satkowski}. In addition, BCPs have many
industrial uses because the length scales involved are smaller or
comparable to the wavelength of light. These applications include
waveguides, photonic band gap materials and other optoelectronic
devices \cite{VBSN01} and dielectric mirrors \cite{FT98}.

Recent studies have highlighted the role of an applied  electric
field in creating well aligned BCP structures. An electric field
has been applied to a polymer film confined by one smooth and one
topographically-patterned electrodes. The field creates an
instability in the polymer film which replicates the pattern on the
electrode \cite{STRS00}. In BCPs, electric field is effective in
aligning micro-domains in a desired direction, as has been shown
theoretically \cite{TAmm02,PW99,AM01} and experimentally
\cite{AH93,AH94}. In a thin film, for example, further removal of
one polymer component can facilitate the creation of
anti-reflection coatings for optical surfaces \cite{WS99} or a
surface with highly ordered features. Lastly, cylindrical domains
of polystyrene (PS)/polymethylmethacrylate (PMMA) diblock
copolymer have been used by Russell, Steiner, Thurn-Albrecht and
co-workers as a basis to an array of long and aligned conducting
domains (nano-wires) with typical size in the range of 100 nm
\cite{RS00}.

The present paper deals with several mechanisms that can be used
to achieve a desired ordering in a BCP melt. We consider in Sec.
2 thin films of A/B BCPs between two flat, parallel surfaces. In the
disordered phase we give a description of the polymer density as
a function of a pre-designed and fixed chemical pattern on the
surface. The decay of surface $q$-modes into the bulk is analyzed
on the level of a linear-response theory. The influence of
confining surfaces on lamellar BCPs is studied in Sec. 3. We find
that for a  one dimensional striped surface pattern (composed of
regions of alternating A and B preference) the lamellae are
tilted
with respect to the parallel surfaces. In this tilted state the
lamellae adjust their periodicity to the surface one, leading to
a better surface coverage. If the surface and lamellar
periodicities are equal, the lamellae are formed perpendicular to
the surfaces. Alignment of confined lamellae by external electric
fields is studied in Sec. 4. It is shown that because different
polymers have different values of the dielectric constant,  the
electrostatic energy favors an orientation of lamellae in a
direction perpendicular to the confining electrodes. This
electrostatic tendency can be used to overcome interfacial
interactions with the bounding electrodes and align structures in
a desired direction.

\section{Confined di-BCP in the disordered phase}

Let us consider first an A/B di-BCP melt in the high temperature
and disordered state, above the Order-Disorder Temperature (ODT)
defined below. The BCP is confined by one or two flat, chemically
patterned surfaces. Although the bulk BCP is disordered above the
ODT, there is an oscillatory decay of the A/B block correlations
and resulting ordering induced by the surface is rather complex.
In the vicinity of the ODT this ordering can become long range
leading to a strong effect. With the definition of the order
parameter $\phi({\bf r})\equiv\phi_A({\bf r})-f$ as the local
deviation of the A monomer concentration from its average, the
bulk free energy can be written as:

\begin{equation}\label{F}
\frac{Nb^3F_b}{k_BT}=\int\left\{\frac12\tau\phi^2+\frac12h\left(
\nabla^2\phi+ q_0^2\phi\right)^2
+\frac16\Lambda\phi^3+\frac{u}{24}\phi^4\right\}{\rm d}^3r
\end{equation}
$d_0=2\pi/q_0$ is the fundamental periodicity in the system, and
is expressed by the polymer radius of gyration $R_g$ through
$q_0\simeq 1.95/R_g$. The parameter $\tau=2\rho
N\left(\chi_c-\chi\right)$  measures the distance from the
critical point ($\tau=0$) in terms of the  Flory parameter
$\chi\sim 1/T$. At the critical point (or equivalently the ODT)
$\chi_c\simeq 10.49/N$. In addition, $b$ is the Kuhn statistical
segment length, $h=1.5\rho c^2R_g^2/q_0^2$ and $\rho=1/Nb^3$ is
the chain density per unit volume. $\Lambda$ and $u$ are the
three- and four-point vertex functions calculated by Leibler
\cite{Leibler80}. Below we restrict ourselves to lamellar phases
of symmetric BCPs ($f=\frac12$, $\Lambda=0$), and set for
convenience $c=u/\rho=1$ throughout the paper.

BCPs \cite{FH87,OK86,TAepl01} and other systems with spatially
modulated phases \cite{Swift77,Andelman95} have been successfully
described by Eq.~(\ref{F}) or similar forms of free energy
functionals. The free energy, Eq.~(\ref{F}), describes  a system
in the disordered phase having a uniform $\phi=0$ for
$\chi<\chi_c$ (positive $\tau$), while  for $\chi>\chi_c$
(negative $\tau$), the system is in the lamellar phase for
$f=\frac12$, $\Lambda=0$, and can be described approximately by a
single $q$-mode $\phi=\phi_L\exp(i{\bf q_0}\cdot {\bf r})$. The
amplitude of the sinusoidal modulations is given by
$\phi_L^2=-8\tau/u$.
The validity of Eq.~(\ref{F}) is limited to a region of the phase
diagram close enough to the critical point where the expansion in
powers of $\phi$ and its derivatives is valid, but not too close
to it, because then critical fluctuations become important
\cite{B-F90,braz1}. This limit employed hereafter is called the
weak segregation limit.

The presence of chemically heterogeneous surfaces is modeled by
adding short-range surface interactions to the free energy of the
form,
\begin{equation}\label{Fs}
F_s=\int\left[\sigma({\bf r_s})\phi({\bf r_s})
+\tau_s\phi^2({\bf
r_s})\right]{\rm d^2}{\bf r_s}
\end{equation}
The integration is carried out over the position of the confining
surfaces parameterized by the vector ${\bf r_s}$. The surface
field $\sigma({\bf r_s})$ has an arbitrary but fixed spatial
variation and is coupled linearly to the BCP surface
concentration
$\phi({\bf r_s})$. Preferential adsorption of the A block
($\phi>0$) onto the {\it entire} surface is modeled by a
constant
$\sigma<0$ surface field, resulting in parallel-oriented layers
(a
perpendicular orientation of the chains).  One way of producing
such a surface field in experiments is to coat the substrate with
random copolymers \cite{L-RPRL96,M-RPRL97}. If the pattern is
spatially modulated, $\sigma({\bf r_s})\neq 0$, then the A and B
blocks are attracted to different regions of the surface. The
coefficient of the $\phi^2$ term in Eq.~(\ref{Fs}) is taken to be
a constant surface correction to the Flory parameter $\chi$
\cite{F87,TAjcp01}. A positive $\tau_s$ coefficient corresponds
to
a suppression of surface segregation of the A and B monomers.

For simplicity we consider first BCP confined by one
surface located at $y=0$ as is depicted in Fig. 1 (a). A
generalization to two parallel surfaces is straightforward
and will be given later. The surface chemical pattern
$\sigma({\bf r}_s)=\sigma(x,z)$ can be decomposed in terms
of its $q$-modes
\begin{equation}
\sigma(x,z)= \sum_{\bf q} \sigma_{\bf q}{\rm e}^{i(q_xx + q_zz)}
\end{equation}
where ${\bf q}=(q_x,q_z)$, and
$\sigma_{\bf q}$ is the mode amplitude. Similarly, $\phi$ can be
written as a sum
\begin{equation}
\phi({\bf r})=\sum_{\bf q}\phi_{\bf q}(y){\rm e}^{i(q_xx+q_zz)}
\end{equation}
Close to the ODT the free energy is stable to second order
in $\phi$, and higher order terms (i.e. the $\phi^4$ term)
can be neglected. Then $\phi({\bf r})$ is inserted into Eq.
(\ref{F}) and an integration over the $x$ and $z$ coordinates
is carried out.
Minimization with respect to $\phi_{\bf q}(y)$ yields the
Euler-Lagrange equation
\begin{equation}\label{EL-fq}
\left[\tau/h+\left(q^2-q_0^2\right)^2\right]\phi_{\bf q}
+2(q_0^2-q^2)\phi_{\bf q}^{\prime\prime}+\phi_{\bf
q}^{\prime\prime\prime\prime}=0
\end{equation}
Note that the equation is linear and that the Fourier harmonics
$\phi_{\bf q}$ are not coupled. The boundary conditions are
rather
complicated because they couple the value of the amplitude and
its
derivatives at the surface. They result from minimization of the
full free energy expression, Eqs.~(\ref{F}) and (\ref{Fs})
\begin{eqnarray}\label{bc1}
\phi_{\bf q}^{\prime\prime}(0)+(q_0^2-q^2) \phi_{\bf q}(0)&=&0\\
 \sigma_{\bf q}/h+2\tau_s\phi_{\bf q}(0)/h+(q_0^2-q^2)\phi_{\bf
q}^{\prime}(0)+\phi_{\bf q}^{\prime\prime\prime}(0)
&=&0\label{bc2}
\end{eqnarray}
Since Eq. (\ref{EL-fq}) is linear, its solution is a sum of
exponentials,
\begin{eqnarray}
\phi_{\bf q}(y)=A_{\bf q}\exp(-k_{\bf
q}y)+B_{\bf q}\exp(-k^*_{\bf q}y)
\end{eqnarray}
where the modulation constant $k_{\bf q}$ and the amplitude
$A_{\bf q}$ are given by
\begin{eqnarray}
k_{\bf q}^2 &=&q^2-q_0^2+i\sqrt{\tau/h}\label{kq}\\ \nonumber
A_{\bf q}&=&-\sigma_q\left(4\tau_s+2{\rm Im}(k_{\bf q})\sqrt{\tau
h}\right)^{-1}
\end{eqnarray}
In the above ${\rm Re}(k_{\bf q})>0$ ensuring that $\phi_{\bf
q}\to 0$ as $y\rightarrow\infty$. This restricts the solution
$\phi_{\bf q}$ to be a sum of only two (out of four) exponential
terms.

The two lengths, $\xi_q=1/{\rm Re}(k_{\bf q})$ and
$\lambda_q=1/{\rm Im}(k_{\bf q})$, correspond to the exponential
decay and oscillation lengths of the ${q}$-modes, respectively.
For fixed $\chi$, $\xi_q$ decreases and $\lambda_q$ increases
with
increasing $q$. Close to the ODT and for  $q>q_0$  we find finite
$\xi_q$ and $\lambda_q\sim(\chi_c-\chi)^{-1/2}$. However, all
$q$-modes in the band $0<q<q_0$ are equally ``active'', i.e.,
these modes decay to zero very slowly in the vicinity of the ODT
as $y\rightarrow\infty$: $\xi_q\sim(\chi_c-\chi)^{-1/2}$ and
$\lambda_q$ is finite. Therefore, the propagation of the surface
imprint (pattern) of $q$-modes with $q<q_0$ into the bulk can
persist to long distances, in contrast to surface patterns with
$q>q_0$ which persist only close to the surface. The $q=q_0$ mode
has both lengths $\xi_q, \lambda_q$ diverging as
$(\chi_c-\chi)^{-1/4}$ for $\chi\rightarrow\chi_c$.

In Fig.~2 we give examples of the polymer morphologies in the
case of three simple surface patterns. A uniform surface [in (a),
$\sigma=\sigma_0$ is constant] causes exponentially decaying
density modulations to propagate in the $y$-direction. A striped
surface [in (b), $\sigma=\sigma_q\cos(qx)$] creates a disturbance
that is periodic in the $x$-direction, while decays exponentially
 in the $y$-direction. The combined surface pattern [in
(c), $\sigma=\sigma_0+\sigma_q\cos(qx)$] induces density
modulations which are the sum of the ones in (a) and (b).

A more complex chemical pattern, shown in Fig.~3 (a), consists of
V shaped stripes on the $y=0$ surface. The polymer density in
parallel planes with increasing distance from the surface is
shown in (b) and (c). Note how the frustration induced by the
tips of surface chemical pattern [in (a)] is relieved as the
distance from the surface increases. Surprisingly, similar
morphology is observed when two grains of lamellar phase meets
with a tilt angle, creating a tilt grain boundary in bulk
systems.

Our treatment of confined BCP can  be easily generalized to the
case of two flat parallel surfaces \cite{TAmm01}. The governing
equation is still Eq. (\ref{EL-fq}), but now there are four
boundary conditions instead of the two in Eqs. (\ref{bc1}) and
(\ref{bc2}). Figure~4 shows how two simple surface patterns can
be
used to achieve a complex three-dimensional polymer morphology,
even though the melt is in its bulk disordered phase. The stripes
on the two surfaces are rotated by 90$^\circ$  with respect to
each other. A symmetric ``checkerboard'' morphology appears in
the mid-plane.

Up to this point, the BCP melt was assumed to be in its bulk
disordered phase (above the bulk ODT point). When a melt in the
lamellar phase (below ODT) is confined in a thin film, the
morphology is dictated by a complex interplay between the natural
periodicity and the imposed film thickness.

\section{Confined lamellar BCP}
In this section we describe the ordering of lamellar BCPs
confined by one or two surfaces. The phase behavior of thin BCP
films in the lamellar phase subject to {\it uniform} surface
fields has been investigated numerically using self-consistent
field (SCF) theory \cite{matsenJCP97,pbmm97} and Monte-Carlo
simulations \cite{G-M-B00,wang1}, and was found to consist of
parallel, perpendicular and mixed lamellar phase denoted
$L_\parallel$, L$_\perp$ and $L_M$, respectively. The latter
$L_M$ phase has parallel lamellae extending from one surface,
which are jointed in a T-junction defect with perpendicular
lamellae extending from the opposite surface
\cite{P-WMM99,wang2}. At a given inter-surface spacing,
increasing the (uniform) surface interactions promotes a parallel
orientation with either A-type or B-type monomers adsorbed onto
the surface. However, if the spacing $L$ between the surfaces is
incommensurate with the lamellar periodicity, or the
incompatibility $\chi$ is increased, a perpendicular orientation
is favored \cite{TAepje01}.

In the treatment given below, a new effect can be observed when
the surfaces are taken to be non-uniform, ``striped'', with
regions of alternating preferences to the A and B blocks [see
Fig.~1 (c)]. The stripe periodicity $d_x$ is assumed to be larger
than the natural (bulk) periodicity, $d_x>d_0$, and the stripes
are modeled by
\begin{equation}\label{sigma2}
\sigma(x,z)=\sigma_q\cos(q_xx)
\end{equation}
and are translational invariant in the $z$-direction. The surface
$q$-mode is $q_x=2\pi/d_x<q_0$.

Contrary to the system above the ODT, a linear response theory
assuming small order parameter as a response to the surface field
is inadequate here, since the bulk phase has an inherent
spatially
varying structure. The surface effects are contained in the
correction to the order parameter
\begin{equation}
\delta\phi({\bf r}) \equiv\phi({\bf r})-\phi_b({\bf r})
\end{equation}
where $\phi_b$ is a ``tilted'' bulk lamellar phase given by
\begin{eqnarray}  \label{bulk}
\phi_b&=&-\phi_L\cos\left(q_xx+q_yy\right)\\
q_x&=&q_0\cos\theta,\qquad q_y=q_0\sin\theta,
\end{eqnarray}
The bulk ordering is depicted schematically as tilted lamellae in
Fig.~1 (c). For the correction order parameter $\delta\phi$ we
choose
\begin{equation}\label{dphi}
\delta\phi(x,y)=g(y)\cos(q_xx).
\end{equation}
This correction describes a lamellar ordering perpendicular to
the surface, and commensurate with its periodicity
$d_x=2\pi/q_x$. The overall morphology of the lamellae is a
superposition of the correction field $\delta\phi$ with the
tilted bulk phase, having a periodicity $d_0$. The region where
the commensurate correction field $\delta\phi$ is important is
dictated by the amplitude function $g(y)$. The total free energy
$F=F_b+F_s$ is now expanded about its bulk value $F[\phi_b]$ to
second order in $\delta \phi$. The variational principle with
respect to $g(y)$ yields a master equation:
\begin{equation}
\label{gov_geqn} \left[ A+C\cos(2q_yy)
\right]g(y)+Bg^{\prime\prime}(y)
+g^{\prime\prime\prime\prime}(y)=0,
\end{equation}
with parameters $A$, $B$ and $C$ given by:
\begin{eqnarray}
A=-\tau/h+q_y^4~,~~~~~B=2q_y^2~,~~~~~ C=-\tau/h~.
\end{eqnarray}
This linear equation for $g(y)$ is similar in form to Eq.
(\ref{EL-fq}) describing the density modulation of a BCP melt in
the disordered phase. The lamellar phase is non-uniform and this
results in a $y$-dependency of the term in square brackets.
The above equation is readily solved using the proper boundary
conditions (for more details see refs. \cite{TAjcp01,TASpre00}).

In Fig.~5 we present results for a BCP melt confined by one
sinusoidally patterned surface,
$\sigma(x)=\sigma_q\cos(q_xx)$, with no
average preference to one of the blocks,
$\langle\sigma\rangle=0$,
for several values of surface periodicity $d_x$ and for fixed
value of the Flory parameter $\chi>\chi_c$. The main
effect of increasing the surface periodicity $d_x$ with respect
to $d_0$ is to stabilize tilted lamellae, with increasing
tilt angle. Note that even for $d_x=d_0$ [Fig.~5a]
yielding no tilt, the perpendicular lamellae have a
different structure close to the surface as is induced by
the surface pattern. Although the surface interactions are
assumed to be strictly local, the connectivity of the
chains causes surface-bound distortions  to propagate into
the bulk of the BCP melt. In particular, this is a strong effect
in the
weak-segregation regime we are considering.

So far in this section we have considered the semi-infinite
problem of a BCP melt confined by one patterned surface. It is of
experimental and theoretical interest to study thin films of BCP
when they are confined between a heterogeneous (patterned)
surface
and a second chemically homogeneous surface. This situation is
encountered when a thin BCP is spread on a patterned surface. The
second interface is the film/air interface and is homogeneous.
Usually the free surface has a lower surface tension with one of
the two blocks. This bias can be modeled by adding a constant
$\sigma_0$ term to the $\sigma(x)$ surface field. For simplicity,
we assume that the surface at $y=-\frac12 L$ has purely
sinusoidal
stripes while at $y=\frac12 L$ the surface is attractive to one
of
the A/B blocks with a constant preference:
\begin{eqnarray}
\sigma(x)&=&\sigma_q\cos(q_xx), \qquad\mbox{at \qquad $y=-\frac12
L$}
,\nonumber\\
\sigma(x)&=&\sigma_0, \qquad\qquad\qquad\mbox{at \qquad
$y=\frac12 L$.}
\end{eqnarray}
A neutral surface at $y=\frac12 L$ is obtained as a special
case with $\sigma_0=0$.
The expression (\ref{bulk}) for the bulk
tilted phase is modified ($y \rightarrow y+\frac12 L$) in order
to match the stripe surface pattern at $y=-\frac12 L$,
\begin{equation}\label{bulk2}
\phi_b=-\phi_L\cos\left[q_xx+q_y(y+\frac12 L)\right]
\end{equation}
The homogeneous surface field at $y=\frac12L$ induces a
lamellar layering parallel to the surface, since the two
A/B blocks are covalently linked together. The simplest
way to account for this layering effect is
 to include an $x$-independent term $w(y)$ in
our ansatz,
Eq.~(\ref{dphi}), for the order parameter:
\begin{equation}\label{dphi2}
\delta\phi(x,y)=g(y)\cos(q_xx)+w(y).
\end{equation}

The tilted lamellar phase confined by one homogeneous and one
patterned surface is a generalization of the mixed (perpendicular
and parallel) lamellar phase, usually referred to as $L_M$. The
latter morphology occurs when the surface imposed periodicity
$d_x$ is equal to the bulk periodicity $d_0$. This ``T-junction''
morphology, shown in Fig.~6, has perpendicular lamellae extending
from the patterned surface. The homogeneous field at the opposite
surface favors a parallel orientation of the lamellae. The
crossover region between the two orientations is found in the
middle of the film, and its morphology depends on temperature
(the
$\chi$ parameter). The effect of the homogeneous field is
evident,
as parallel ordering extends from the top surface. We see here
that strong enough modulated surface fields stabilize the tilted
lamellar phases and, in particular, the $L_M$ phase.

In the discussion above we have considered ordering mechanisms
where the interaction of the polymers with the confining surfaces
is mediated to regions far from the surfaces because of chain
connectivity. We now turn to discuss orientation of BCP films in
presence of external electric fields. This is a bulk ordering
mechanism that does not originate from surface interactions.

\section{Alignment by electric fields}

A well known mechanism to cause orientation or
structural changes in polarized media is the ``dielectric
mechanism''. This effect is based on the fact that when a
material with inhomogeneous dielectric constant is placed
in an electric field $E$, there is an electrostatic free
energy penalty for having dielectric interfaces
perpendicular to the field \cite{TAmm02,Onuki95,TDR00}.
Thus, a state where $\nabla\eps$ is perpendicular to the field
${\bf E}$ is favored. For layered materials such
as lamellar BCP phases, the strength of this effect is
proportional to $(\eps_A-\eps_B)^2E^2$, where $\eps_A$ and
$\eps_B$ are the dielectric constants of the two
micro-domains, and is enhanced when the difference in
polarizabilities
is large.

Let us first consider lamellae confined in a thin film with no
electric field. We will then include an electric field and see its
influence.  For simplicity and brevity of presentation we assume
that the Flory parameter $\chi$ does not change on the surface,
$\tau_s=0$. We also concentrate on uniform surface fields with
equal magnitudes, $\sigma(y=-\frac12 L)=\pm\sigma$,
$\sigma(y=\frac12 L)=\pm\sigma$. If the surface affinity $\sigma$
is sufficiently large, the lamellae will order in a parallel
arrangement as was discussed in the previous sections. These
lamellae stretch or compress, increasing the bulk free energy, in
order to decrease their surface energy. We mention below an
adaptation \cite{TAepje01} to the strong stretching approximation
of Turner \cite{8:turnerPRL92} and Walton {\it et al.}
\cite{8:W-RMM94}. The lamellar period is $d_0=2\pi/q_0$ and $m$
is the closest integer to $L/d_0$. Depending on the value of
$\sigma$, an integer ($n=m$) or half integer ($n=m+\frac12$)
number of lamellae fill the gap between the two surfaces. In the
former case the ordering is symmetric (the same type of monomers
adsorbed onto both surfaces), while in the latter it is asymmetric
(A monomers adsorbed on one surface, whereas B monomers adsorbed
on the other surface). The parallel lamellae are described by an
order parameter $\phi_\parallel$ given by \cite{TAepje01}
\begin{equation}\label{8:phipar}
\phi_\parallel(x,y)=\pm\phi_L\cos[q_{_\parallel}(y+\frac12 L)]\\
\end{equation}
The wavenumber is $q_{_\parallel}=2\pi n/L$, and the choice of
$\pm$ signs in Eq. (\ref{8:phipar}) is such that the surface
interactions, Eq. (\ref{Fs}), are minimized. The perpendicular
lamellae have the unperturbed bulk periodicity
$d_0=2\pi/q_0$ and their order parameter is simply given by
\begin{equation}\label{8:phiperp}
\phi_\perp(x,y)=\phi_L\cos(q_0 x)\\
\end{equation}

We consider now the case where an electric field is turned
on, in a direction that is perpendicular to the surfaces.
Under conditions of constant voltage difference across the
electrodes situated at the two bounding surfaces, the
minimum of the free energy is obtained by maximizing the
capacitance. Since the A- and B-monomers have
different dielectric constants, the effect of the electric
field is to align the BCP layers parallel to the field,
i.e. perpendicular to the surfaces. At a certain field
strength this tendency starts to dominate over
the preference induced by the surfaces to have
parallel lamellae. Further
increase of $E$ above the threshold value gives rise to a
perpendicular lamellar ordering.

In the weak segregation regime,
the small value of density modulations $\phi$ has been used
\cite{TAmm02,AH93,fraaije02}
in order to
write the electrostatic free energy as
\begin{eqnarray}\label{8:Fel}
\frac{Nb^3F_{\rm el}}{k_BT}=\beta\int (\hat{{\bf q}}\cdot{\bf
E})^2\phi_{\bf q}\phi_{-\bf q}{\rm
d}^3q\\
\beta=\frac{(\eps_A-\eps_B)^2Nb^3}{4(2\pi)^4
k_BT\langle\eps\rangle}\label{8:beta}
\end{eqnarray}
Here $\phi_{\bf q}$ is the Fourier transform of $\phi({\bf
r})$: $\phi({\bf r})=\int \phi_{\bf q} {\rm exp}(i{\bf
q\cdot r}){\rm d}{\bf q}$, and ${\bf \hat{q}}={\bf q}/q$
is a unit vector in the ${\bf q}$-direction. Copolymer
modulations with a non-vanishing component of the
wavenumber ${\bf q}$ along the electric field, have a
positive contribution to the free energy. In other words,
there is a free energy penalty for having dielectric
interfaces in a direction perpendicular to the electric
field. In Eq. (\ref{8:beta}), $\eps_A$ and $\eps_B$ are
the dielectric constants of the pure A and B-blocks,
respectively \cite{AH93}, and
$\langle\eps\rangle=\frac12(\eps_A+\eps_B)$ is the average
dielectric constant in the symmetric BCP film ($f=\frac12$). In
the weak
segregation regime, the spatial variations in $\phi$ are
small and $\eps$ varies linearly with $\phi$,
\begin{eqnarray}
\eps(\phi)&\simeq&(\frac12+\phi)\eps_A+(\frac12-\phi)\eps_B
\nonumber\\&=&\langle\eps\rangle+(\eps_A-\eps_B)\phi
\end{eqnarray}
The total free energies of the parallel and perpendicular
lamellae
per unit area of the film are then given by
\begin{eqnarray}
F_\parallel&=&\frac14\left[\tau+h\left(q_0^2-
q_{_\parallel}^2\right)^2+2\beta
E^2+\frac{u}{16}\phi_L^2\right]\phi_L^2L+\Sigma\\
F_\perp&=&\frac14\left[\tau+
\frac{u}{16}\phi_L^2\right]\phi_L^2L
\end{eqnarray}
$\Sigma=(\pm\phi_L\pm \phi_L)\sigma$ is the total  surface
interaction per unit area [Eq. (\ref{Fs})] at the two walls,  and
is negative.
The
$\pm$ sign is determined from the $\pm$ sign of the order
parameter in Eq. (\ref{8:phipar}). When the two surfaces are
chemically identical, two similar signs in $\Sigma$
mean symmetric ordering with an integer number of lamellae
between the
electrodes ($\Sigma=\pm 2\phi_L\sigma$). In an antisymmetric
ordering with a half-integer
number of confined lamellae, $\Sigma=0$ as the two wall
preferences cancel each other.

In Fig.~7 we compare the relative stability of parallel
and perpendicular lamellae as a function of surface interaction
strength $\sigma$ (on the vertical axis) and surface separation
$L$
(on the horizontal axis). In (a) the electric field is zero.
Parallel
lamellae are favored when surface separation $L$ is close
to an integer value (recall that we take in this example two
surfaces
with the same polymer affinity $\sigma$). Perpendicular
ordering is favored when lamellae frustration is largest,
i.e. when $L/d_0$ is approximately a half-integer number.
A degeneracy between the two states occurs for these
values of $L$ when the surface interactions vanish,
$\sigma=0$. For more details and a generalization to
surfaces with different affinity for the A and B blocks see
ref. \cite{TAepje01}. In (b) an electric field is applied
in a direction perpendicular to the electrodes. As a
consequence, perpendicular lamellae are favored and the boundary
line between
the
light and dark regions in Fig.~7 is shifted upward to higher
$\sigma$ values.

In the strong segregation regime ($\chi\gg\chi_c$), thin BCP films
subjected to external electric field can be analyzed as well.
Since in this regime the correlation length is not larger than
the system size (in contrast to the weak segregation limit
mentioned above), another {\it finite} length scale enters into
the problem. The result is that the system has three possible
states: parallel state in which parallel lamellae span the whole
film, a perpendicular state, and a mixed state with parallel
lamellae near the surfaces and perpendicular lamellae in the
middle of the film. These three states are separated by three
critical fields $E_1$, $E_2$ and $E_3$.

In Fig. 8 (a) we show the phase diagram in the $\delta$--$E$
plane,
where $\delta$ is the dimensionless difference between then A and
B-block surface energies and $E$ is the strength of applied
electric
field. Transition from parallel to mixed lamellae at $E=E_1$
occurs if $\delta$ is above some threshold value,
$\delta^*$. It is then followed by a transition between the mixed and
perpendicular states at $E=E_2$. If $\delta$ is small,
$\delta<\delta^*$, there is a direct transition from parallel to
perpendicular lamellae at $E=E_3$.

A different cut through the phase diagram in the  $L$--$E$
(while keeping $\delta$ constant)
is shown
in Fig. 8 (b). A transition between
parallel and mixed states occurs at $E=E_1$, followed by
a second transition at $E=E_2$ from the mixed to the
perpendicular
state, provided that surface
separation is large enough, $L>L^*$. When $L<L^*$ a direct
transition between parallel and perpendicular lamellae occurs at
$E=E_3$.

\section{Summary}

Several ordering mechanisms in confined BCPs are considered in
this paper. Above the order-disorder temperature the polymer
density near a chemically patterned surface is given as a
function of the surface pattern. It is shown that each surface
$q$-mode of the chemical pattern arises its own density mode, and
that these modes can be regarded as uncoupled (linear response
theory). The decay of these density modes into the bulk (away
from the surfaces) is analyzed. In the weak segregation regime
employed here, surface correlations become long range and hence
simple chemical patterns (as in Fig.~4) can yield complex
copolymer morphology, even though the bulk  is in its disordered
phase.

We describe lamellar BCP when they are confined by striped
surface whose periodicity is larger than the lamellar
periodicity. We find that in this case lamellae will tend
to tilt with respect to the surface in order to
optimize their surface interactions. The deviations from the
perfect lamellar shape, induced by the surfaces, are
obtained. These undulations of the lamellar interface are
more prominent as the ODT is approached. Mixed lamellar
phases appear when one surface has chemically patterns in the
form of stripes while
 the
other is uniform.

We examine the effect of an external electric field on the phase
behavior of parallel and perpendicular lamellae. In the absence of
any external electric field, we recover the previously obtained
phase diagram \cite{matsenJCP97,pbmm97,G-M-B00}. The influence of
electric field [Fig.~7 (b)] is to favor the perpendicular
lamellae: the region where parallel lamellae are stable is pushed
up-wards in the phase diagram. We present as well the phase
diagrams in the strong segregation regime. In this regime there
are three possible system states (parallel, perpendicular and
mixed) with three critical fields which separate them, as can be
seen in Figure~8 (a) and (b).

The analytical calculations presented here together with other
experimental and numerical studies are useful tools towards
obtaining well controlled structures in the nanoscale. Indeed, the
technical details employed should not obscure some simple results
which are rather universal: the use of two simple one-dimensional
striped surface patterns in order to achieve a complex
three-dimensional morphology [Fig.~4], and  surface field
periodicity that facilitates control of tilt angle [Fig.~5].
Finally, tuning the strength of electric field and surface
separation (or interaction) as an external perturbation to create
parallel, mixed or perpendicular lamellae [Fig.~8]. These
mechanisms call for further application-oriented studies as well
as theoretical ones.

\bigskip
{\it Acknowledgments}~~~ Partial support from the U.S.-Israel
Binational Foundation (B.S.F.) under grant No. 98-00429 and the
Israel Science Foundation founded by the Israel Academy of
Sciences and Humanities
--- centers of Excellence Program is gratefully
acknowledged. One of us (Y.T.) acknowledges support from the
Chateaubriand Fellowship Program.


\newpage

\newpage
\bigskip
{\bf \Large{Figure Captions}}

\begin{itemize}

\item{{\bf Figure 1.} Schematic illustration of
the coordinate system for BCP confined by one [part (a)] or two
[part (b)] planar and parallel surfaces. (c) Lamellae are formed
tilted with respect to the surface if the surface periodicity
$d_x$ is larger than the natural one $d_0$.}

\item{{\bf Figure 2.} A BCP melt confined by one
surface at $y=0$. B-monomer density is high in dark regions,
while
A monomers are in light regions. In (a) the surface is uniform,
$\sigma=0.3$ and in (b) it has stripes given by
$\sigma=0.3\cos(\frac23q_0)$. The ``combined'' effect is shown in
part (c) where $\sigma=0.3+0.3\cos(\frac23q_0x)$ has a uniform
and
modulated part. The Flory parameter is $N\chi=10.2$, $\tau_s=0$
and lengths in the $x$ and $z$ directions are scaled by the
lamellar period $d_0$. }

\item{{\bf Figure 3.} Propagation of surface
pattern into the bulk. The surface pattern in the $y=0$ plane is
shown in (a), where white (black) show regions preferring A (B)
monomers. Parts (b) and (c) are contour plots of the polymer
density at $y=3d_0$ and $y=8d_0$, respectively. $N\chi=9.5$ and
$\tau_s=0$. }

\item{{\bf Figure 4.} BCP melt confined by two
flat parallel striped surfaces, depicted in parts (a) and (c),
and
located at $y=-d_0$ and $y=d_0$, respectively. The melt
morphology
in the mid-plane ($y=0$) is shown in part (c). The Flory
parameter
is $N\chi=9$ and $\tau_s=0$. }

\item{{\bf Figure 5.} Tilted lamellar phase
between two parallel patterned surfaces close to a patterned
bounding surface. (See also Fig. 1c). The surface patterning is
modelled by the term $\sigma_q\cos(2\pi x/d_x)$. The lamellae tilt
angle $\theta=\arccos(d_0/d_x)$ increases as the periodicity of
the surface $d_x$ increases: $\theta=0$ for $d_x=d_0$ in (a),
$\theta\simeq 48.1^o$ for $d_x=\frac32 d_0$ in (b) and
$\theta\approx 70.5^o$ for $d_x=3 d_0$ in (c). In the plots
$\sigma_q/hq_0^3\phi_L=1$. The Flory parameter $N\chi=11.5$ and
$\tau_s/hq_0^3=0.1$.}

\item{{\bf Figure 6.} A BCP confined film
showing a crossover from perpendicular lamellae at the
$y=-\frac12
L=-2d_0$ surface to parallel lamellae at the other surface,
$y=\frac12 L$. The pattern on the bottom surface,
$\sigma(x)=\sigma_q\cos(q_0x)$, has the bulk periodicity $d_0$,
and amplitude $\sigma_q=2hq_0^3$, while the top surface
($y=\frac12 L$) is homogeneously attractive to the B polymer (in
black), $\sigma_0=4hq_0^3$. The Flory parameter is given by
$N\chi=10.7$ and $\tau_s/hq_0^3=0.4$. }

\item{{\bf Figure 7.} The stability of
$L_\parallel$ (in light) vs. $L_\perp$ lamellae (dark), as a
function of wall separation $L$ and interfacial interaction
parameter $\sigma$. The latter is taken as $\sigma>0$ on both
surfaces. In (a) the electric field
strength is zero, while in (b) $E\sqrt{\beta}=0.02$. The
$L_\parallel$ phase is pushed upward in the stability diagram in
(b), removing the degeneracy between $L_\perp$ and $L_\parallel$
that occurs for neutral walls ($\sigma=0$) for $L/d_0$ being an
integer number. The Flory parameter is $\chi N=11$.}

\item{\bf Figure~8.} (a) Phase diagram in the $E$-$\delta$ plane.
$\delta$ is the difference between the A and B-block surface
energies. When $\delta<\delta^*$, there is a direct transition
between parallel and perpendicular lamellae at $E=E_3$. For
$\delta>\delta^*$, a transition between parallel and mixed states
occurs at $E=E_1$, followed by a second transition between mixed
and perpendicular states when $E=E_2>E_1$. (b) Similar diagram,
but in the $E$-$L$ plane, where $L$ is surface separation $d_0$
is
the lamellar period and $\delta=5$.

\end{itemize}
\newpage

\begin{figure}[h!]
\begin{center}
\includegraphics[scale=0.9,bb=25 405 570 640,clip]{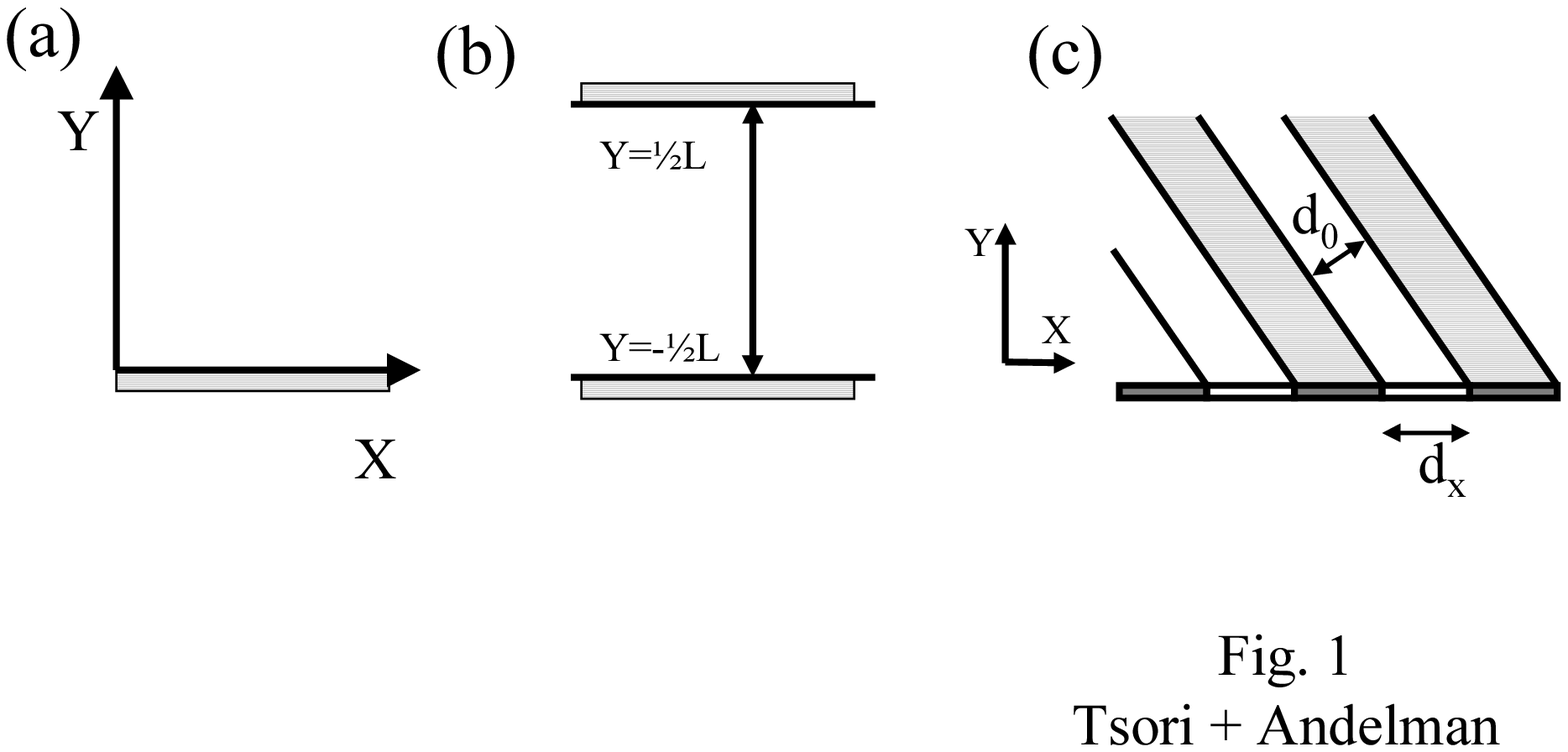}
\end{center}
\end{figure}

\begin{figure}[h!]
\begin{center}
\includegraphics[scale=0.9,bb=15 240 580 475,clip]{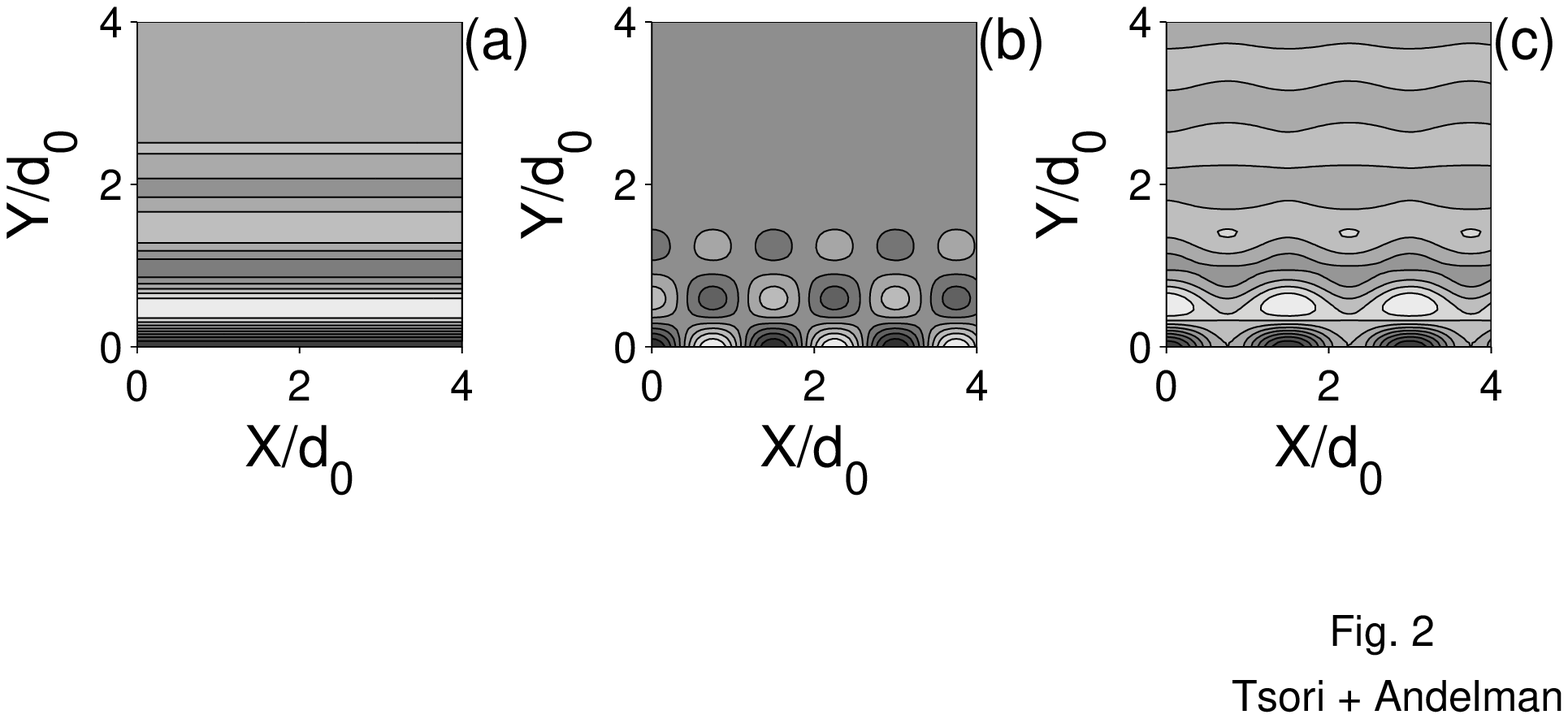}
\end{center}
\end{figure}

\begin{figure}[h!]
\begin{center}
\includegraphics[scale=0.9,bb=15 248 550 490,clip]{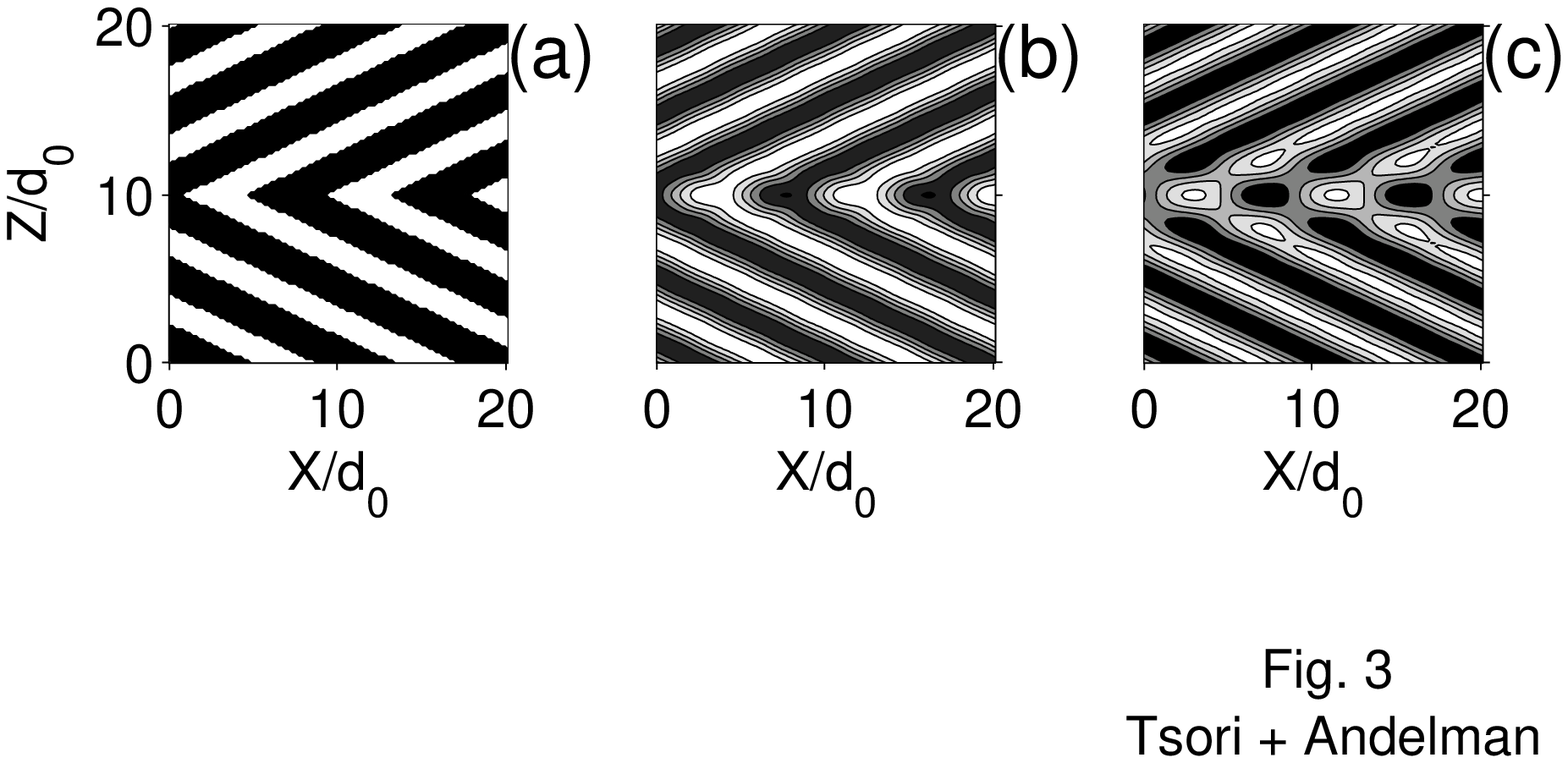}
\end{center}
\end{figure}

\begin{figure}[h!]
\begin{center}
\includegraphics[scale=0.95,bb=45 229 570 490,clip]{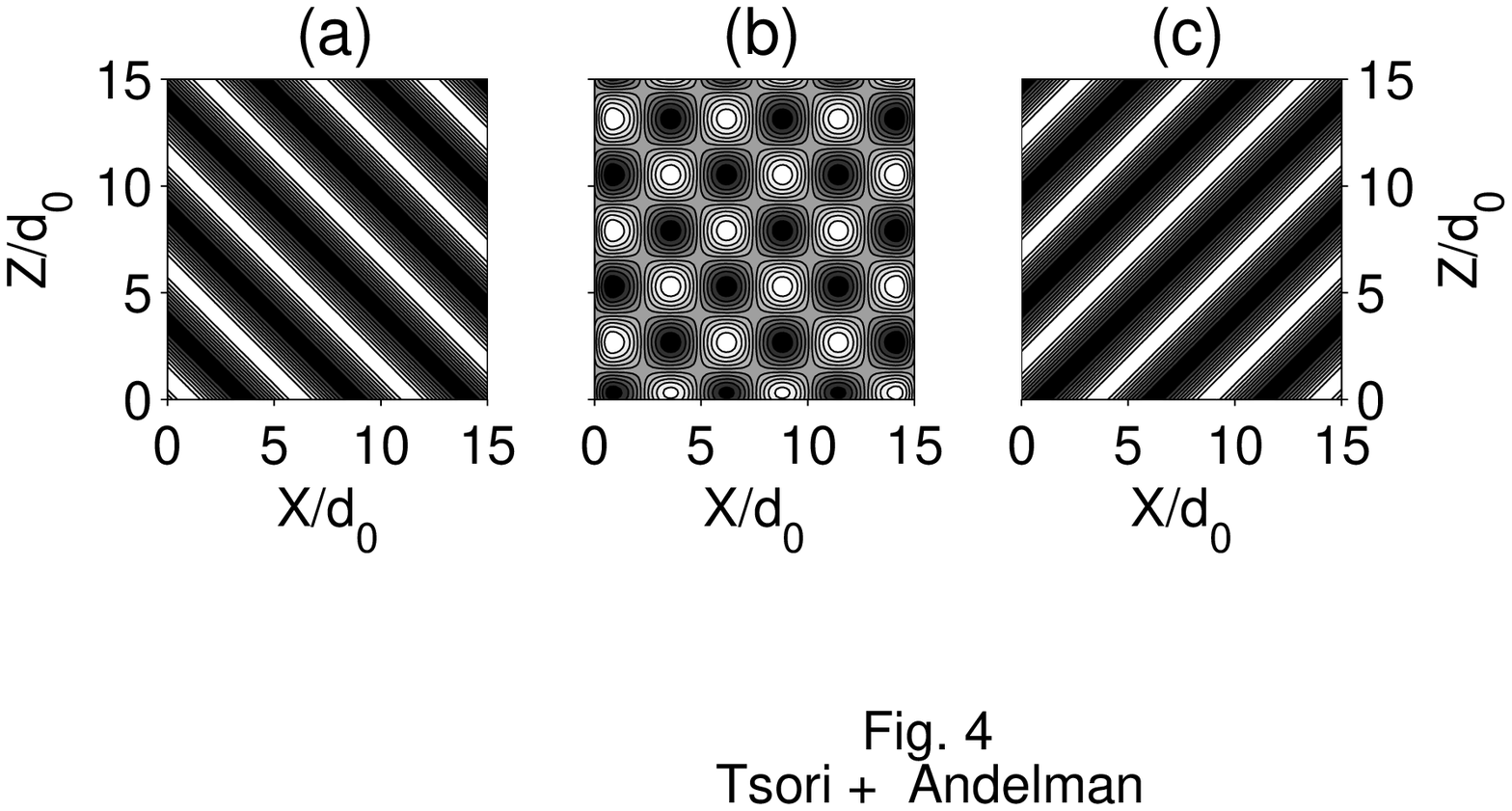}
\end{center}
\end{figure}

\begin{figure}[h!]
\begin{center}
\includegraphics[scale=0.9,bb=20 273 560 485,clip]{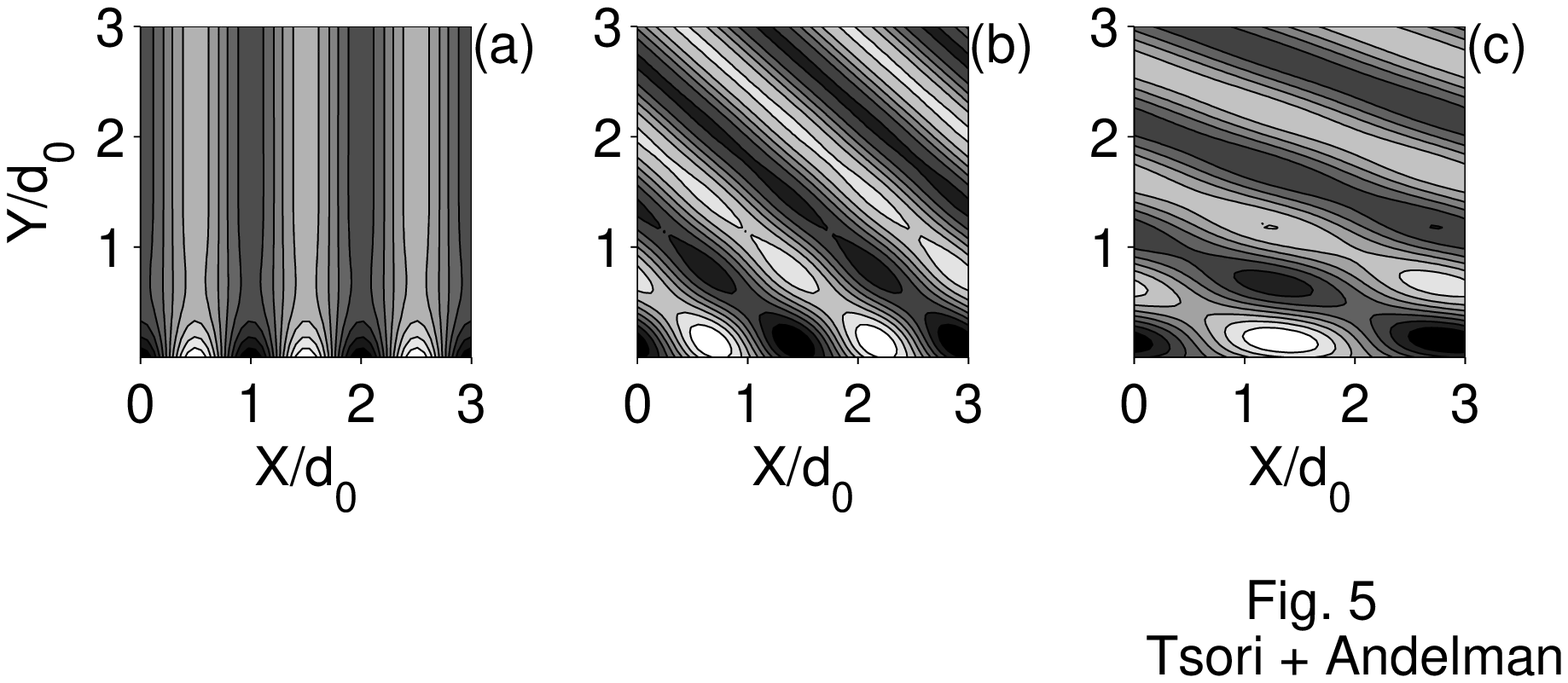}
\end{center}
\end{figure}

\begin{figure}[h!]
\begin{center}
\includegraphics[scale=0.7,bb=50 163 580 680,clip]{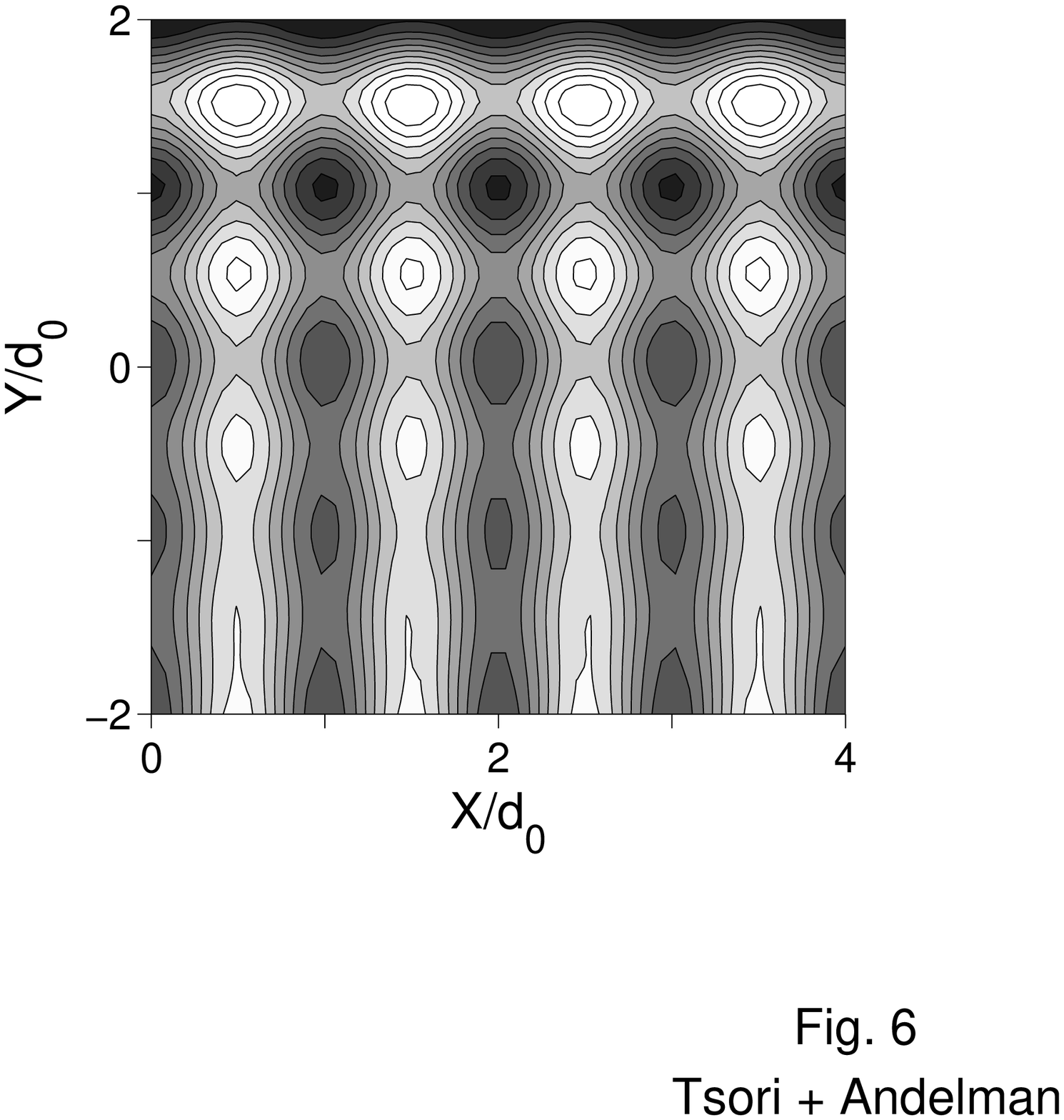}
\end{center}
\end{figure}

\begin{figure}[h!]
\begin{center}
\includegraphics[scale=0.8,bb=15 247 580 535,clip]{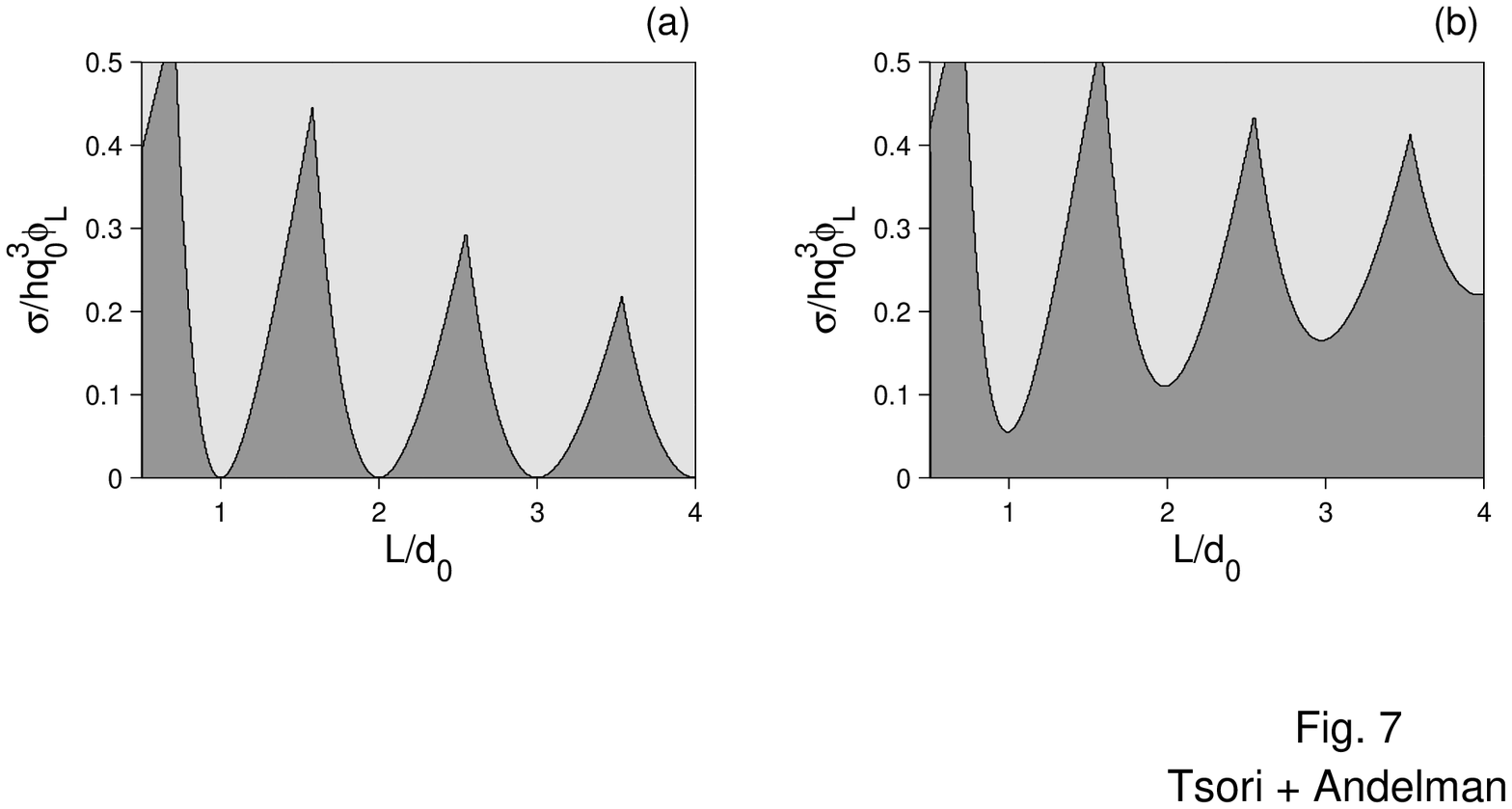}
\end{center}
\end{figure}

\begin{figure}[h!]
\begin{center}
\includegraphics[scale=1.2,bb=195 202 400 610,clip]{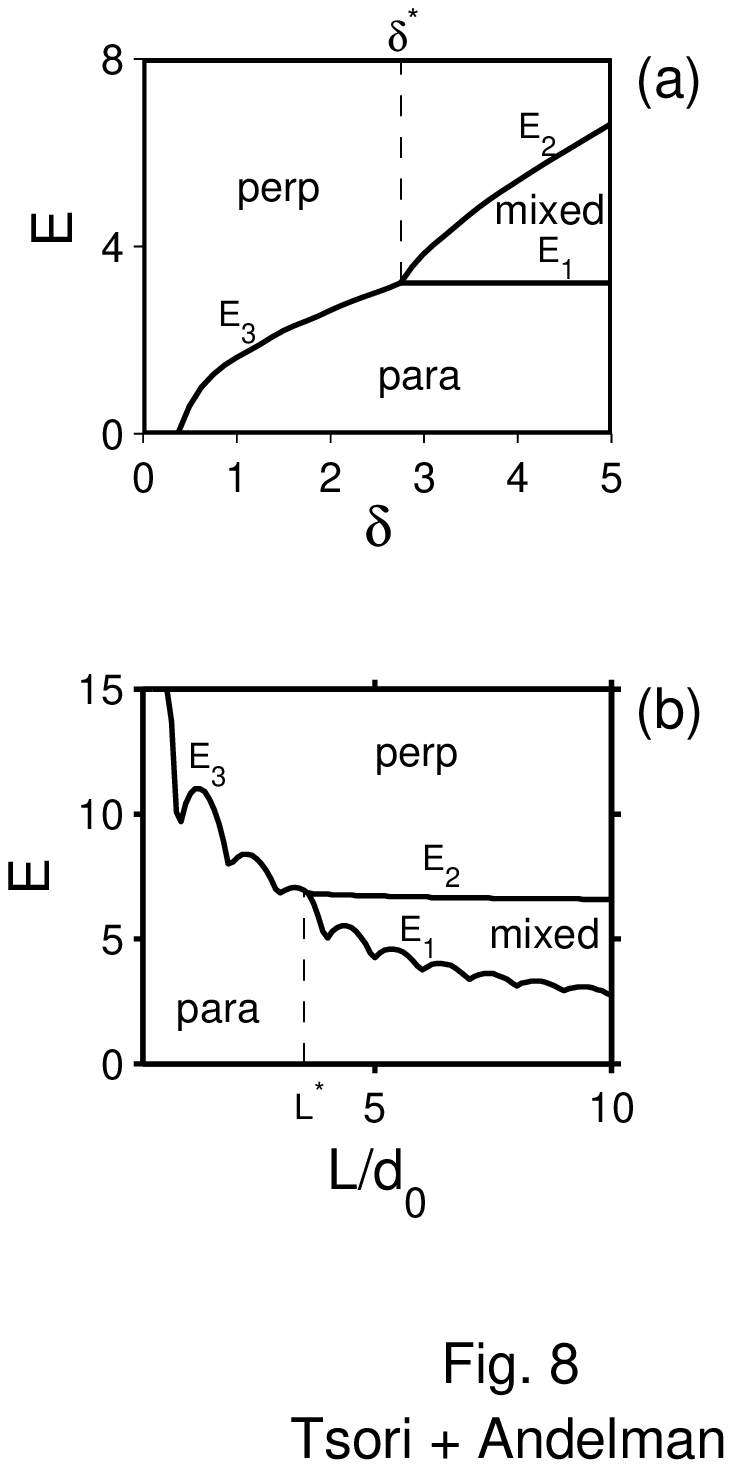}
\end{center}
\end{figure}


\begin{thebibliography}{0}

\bibitem{hamley1} I. W. Hamley, {\it The Physics of Block
Copolymers}, Oxford University:
Oxford, 1998.

\bibitem{yachin1} Y. Cohen, R. J. Albalak, B. J. Dair, M. S.
Capel
and E. L. Thomas, {\it Macromolecules} {\bf 33}, 6502 (2000).

\bibitem{yachin2} Y. Cohen,
M. Brinkmann and E. L. Thomas, {\it J. Chem. Phys.} {\bf 114},
984 (2001).

\bibitem{satkowski} Z.-R. Chen, J. A. Kornfield, S. D. Smith, J.
T.
Grothaus and M. M. Satkowski, {\it Science} {\bf 277}, 1248
(1997).

\bibitem{VBSN01} Y. A. Vlasov, X. Z. Bo, J. C. Sturm and D. J.
Norris,
{\it Nature} {\bf 414}, 289 (2001).

\bibitem{FT98} Y. Fink, J. N. Winn, S. Fan, C. Chen, J.
Michel, J. D. Joannopoulos and E. L. Thomas {\it Science}
{\bf 282}, 1679 (1998).

\bibitem{STRS00}  E. Sch\"{a}effer, T.  Thurn-Albrecht, T. P.
Russell and
U.
Steiner, {\it Nature} {\bf 403}, 874 (2000).

\bibitem{TAmm02} Y. Tsori and D. Andelman, {\it
Macromolecules} {\bf 35}, 5161 (2002).

\bibitem{PW99} G. G. Pereira and D. R. M. Williams, {\it
Macromolecules} {\bf 32}, 8115 (1999).

\bibitem{AM01} B. Ashok, M. Muthukumar and T. P.
Russell, {\it J. Chem. Phys.} {\bf 115}, 1559 (2001).

\bibitem{AH93} K. Amundson, E. Helfand, X. Quan and
S. D. Hudson, {\it Macromolecules} {\bf 26}, 2698 (1993).

\bibitem{AH94} K. Amundson, E. Helfand and X. Quan, {\it
Macromolecules} {\bf 27}, 6559 (1994).

\bibitem{WS99} S. Walheim,  E. Sch\"{a}ffer, J.
Mlynek and U. Steiner,  {\it Science} {\bf 283}, 520
(1999).

\bibitem{RS00} T. Thurn-Albrecht, J. Schotter, G. A. K\"{a}astle,
N. Emley, T. Shibauchi, L. Krusin-Elbaum, K. Guarini, C. T.
Black,
M. T. Tuominen and T. P. Russell, {\it Science} {\bf 290}, 2126
(2000).

\bibitem{Leibler80} L. Leibler, {\it Macromolecules}
{\bf 13}, 1602 (1980).

\bibitem{FH87}  G. H. Fredrickson and E. Helfand, {\it J. Chem.
Phys.}
{\bf 87}, 697 (1987).

\bibitem{OK86} K. Ohta and K. Kawasaki, {\it Macromolecules}
{\bf 19}, 2621, (1986).

\bibitem{TAepl01} Y. Tsori and D. Andelman, {\it Europhys.
Lett.} {\bf 53}, 722 (2001).

\bibitem{Swift77} J. Swift and P. C. Hohenberg, {\it Phys. Rev.
A} {\bf 15}, 319 (1977).

\bibitem{Andelman95} M. Seul and D. Andelman, {\it Science}
{\bf 267}, 476 (1995).



\bibitem{B-F90} F. S. Bates and G. H. Fredrickson,
{\it Annu. Rev. Phys. Chem.} {\bf 41}, 525 (1990).

\bibitem{braz1} S. A. Brazovskii, {\it
Sov. Phys. JETP} {\bf 41}, 85 (1975).

\bibitem{L-RPRL96} G. J. Kellogg, D. G. Walton, A. M. Mayes A.
M.,
P. Lambooy, T. P. Russell, P. D. Gallagher and S. K. Satija,
{\it Phys. Rev. Lett.} {\bf 76}, 2503 (1996).

\bibitem{M-RPRL97} P. Mansky, T. P. Russell, C. J. Hawker, J.
Mayes,
D. C. Cook and S. K. Satija, {\it Phys. Rev. Lett.} {\bf 79}, 237
(1997).

\bibitem{F87} G. H. Fredrickson, {\it Macromolecules} {\bf 20},
2535
(1987).

\bibitem{TAjcp01} Y. Tsori and D. Andelman, {\it J. Chem. Phys.}
{\bf
115}, 1970 (2001).

\bibitem{TASpre00} Y. Tsori, D. Andelman and M. Schick,
{\it Phys. Rev. E.} {\bf 61}, 2848 (2000).

\bibitem{TAmm01} Y. Tsori and D. Andelman,
{\it Macromolecules} {\bf 34}, 2719 (2001).


\bibitem{matsenJCP97} M. W. Matsen, {\it J. Chem. Phys.} {\bf
106},
  7781 (1997).

\bibitem{pbmm97} G. T. Pickett and A. C. Balazs, {\it
Macromolecules}
{\bf 30}, 3097 (1997).

\bibitem{G-M-B00} T. Geisinger, M. Mueller and K. Binder, {\it J.
Chem. Phys.} {\bf 111}, 5241 (1999).

\bibitem{wang1} Q. Wang, Q. Yan, P. F. Nealey and J. J. de Pablo,
 {\it J. Chem. Phys.} {\bf 112}, 450 (2000).

\bibitem{P-WMM99} G. G. Pereira and D. R. M. Williams,
{\it Macromolecules} {\bf 32}, 1661 (1999); {\it ibid}, 758
(1999).

\bibitem{wang2} Q. Wang, S. K. Nath, M. D. Graham, P. F. Nealey
and J. J. de Pablo, {\it J. Chem. Phys.} {\bf 112}, 9996 (2000).

\bibitem{Onuki95} A. Onuki and J. Fukuda, {\it Macromolecules}
{\bf 28}, 8788 (1995).

\bibitem{TDR00} T. Thurn-Albrecht, J. DeRouchey and T. P.
Russell, {\it Macromolecules} {\bf 33}, 3250 (2000).

\bibitem{TAepje01} Y. Tsori and D. Andelman, {\it Eur. Phys.
J. E} {\bf 5}, 605 (2001).

\bibitem{8:turnerPRL92} M. S. Turner, {\it Phys. Rev. Lett.}
{\bf69}, 1788 (1992).

\bibitem{8:W-RMM94} D. G. Walton, G. J. Kellogg, A. M. Mayes,
P. Lambooy and T. P. Russell, {\it Macromolecules} {\bf
27}, 6225 (1994).


\bibitem{fraaije02} A. V. Kyrylyuk, A. V. Zvelindovsky,
G. J. A. Sevink and J. G. E. M. Fraaije, {\it
Macromolecules} {\bf 35}, 1473 (2002).


\end{thebibliography}
\end{document}